\documentclass[letter]{aa} 
\usepackage[varg]{txfonts}
\usepackage{graphicx}
\usepackage{txfonts}
\usepackage{natbib}
\bibpunct{(}{)}{;}{a}{}{,} 

\begin{document}

\title{Main sequence dynamo magnetic fields emerging in the white dwarf phase}

\author{Maria Camisassa\inst{1},
    J. R. Fuentes \inst{2}, Matthias R. Schreiber \inst{3},
    Alberto Rebassa-Mansergas\inst{1,4},
    Santiago Torres \inst{1,4},
    Roberto Raddi\inst{1}, 
    \and      
    Inma Dominguez \inst{5}
    } 
\institute{Departament de F\'\i sica, 
           Universitat Polit\`ecnica de Catalunya, 
           c/Esteve Terrades 5, 
           08860 Castelldefels, 
           Spain
           \and
            Department of Applied Mathematics, University of Colorado Boulder, Boulder, CO 80309-0526, USA
     	    \and
            Departamento de Física, Universidad Técnica Federico Santa María, Av. España 1680, Valparaíso, Chile
          \and
           Institute for Space Studies of Catalonia, 
           c/Gran Capita 2--4, Edif. Nexus 201, 
           08034 Barcelona,  Spain
            \and
            Departamento de Fısica Teorica y del Cosmos, Universidad de Granada, 18071 Granada, Spain
           \\
          }

\date{Received ; accepted }

\abstract{Recent observations of volume-limited samples of magnetic white dwarfs (WD) have revealed a higher incidence of magnetism in older WDs. Specifically, these studies indicate that magnetism is more prevalent in WDs with fully or partially crystallized cores compared to those with entirely liquid cores. This has led to the recognition of a crystallization-driven dynamo as an important mechanism for explaining magnetism in isolated WDs. However, recent simulations {challenged} the capability of this mechanism to generate surface magnetic fields with the typical strengths detected in WDs.
In this letter, we explore an alternative hypothesis for the surface emergence of magnetic fields in isolated WDs. WDs with masses $\gtrsim 0.55 M_\odot$ are the descendants of main-sequence stars with convective cores capable of generating strong dynamo magnetic fields. This idea is supported by asteroseismic evidence of strong magnetic fields buried within the interiors of red giant branch stars.
Assuming that these fields are disrupted by subsequent convective zones, we have estimated magnetic breakout times for WDs with carbon-oxygen (CO) cores and masses ranging from $0.57 M_\odot$ to $1.3 M_\odot$.
Due to the significant uncertainties in breakout times stemming from the treatment of convective boundaries and mass loss rates, we cannot provide a precise prediction for the emergence time of the main-sequence dynamo field. However, we can predict that this emergence should occur during the WD phase for WDs with masses $\gtrsim 0.65 M_\odot$. We also find that the magnetic breakout is expected to occur earlier in more massive WDs, consistently with observations from volume-limited samples and the well-established fact that magnetic WDs tend to be more massive than non-magnetic ones.
Moreover, within the uncertainties of stellar evolutionary models, we find that the emergence of main-sequence dynamo magnetic fields can account for a significant portion of the magnetic WDs. Additionally, we estimated magnetic breakout times due to crystallization-driven dynamos in CO WDs suggesting that this mechanism cannot explain the majority of magnetic WDs.
}
\titlerunning{Magnetic fields emerging in the white dwarf phase}
\authorrunning{Camisassa et al.}  

\maketitle


\section{Introduction}
\label{introduction}

 WD stars are the most common end point of stellar evolution as all main sequence stars with masses lower than $9-12M_{\odot}$ will eventually become WDs. Therefore, the WD population is considered a powerful tool to investigate a wide variety of astrophysical problems, from the formation and evolution of our Galaxy to the ultimate fate of planetary systems \citep[see][for a review]{2010A&ARv..18..471A}. In particular, WDs contain valuable information about the evolution of their progenitor stars, and can be used to constrain nuclear reaction rates \citep{2019A&A...630A.100D}, the initial-to-final-mass relation \citep{2008MNRAS.387.1693C,2018ApJ...866...21C} and the occurrence of third dredge-up episodes in the Asymptotic Giant Branch (AGB) \citep{2015A&A...576A...9A}, among others.

The presence of magnetic fields in the surface of WDs has been known for more than 50 years \citep{1970ApJ...161L..77K,1970ApJ...160L.147A}, yet its origin is still not well understood \citep[see][for a review]{2015SSRv..191..111F}. Several explanations have been proposed, involving both single evolution and binary interactions. One possibility is that the WDs inherited a magnetic field from their formation history, being this consistent with the magnetic fields observed in the surface of peculiar Ap and Bp stars. More recently, it has been proposed that the mixing instability induced by WD crystallization in fast-rotating WDs can generate a dynamo magnetic field \citep{2017ApJ...836L..28I}. Other explanations involve close binary interactions: either the magnetic field can be generated during a merger episode \citep{2012ApJ...749...25G}, or in a dynamo acting during a post-common-envelope phase \citep{2008MNRAS.387..897T}. Although there are more than 600 magnetic WDs reported in the literature \citep{2020AdSpR..66.1025F}, an important step forward in the interpretation of the origin of WD magnetism comes from the recent determination of a 20pc volume limited sample of magnetic WDs \citep{2021MNRAS.507.5902B,2007ApJ...654..499K}. These authors have checked individually each of the WDs within 20 pc from the Sun for the presence of magnetic fields, 
thus eliminating the observational biases of the previous magnitude limited samples of magnetic WDs. Based on their analysis, \cite{2021MNRAS.507.5902B} conclude that the occurrence of magnetism is significantly higher in WDs that have undergone the process of core crystallization than in WDs with fully liquid cores. \cite{2022ApJ...935L..12B},
expanded the volume limited sample to 40 pc, although only including WDs younger than 0.6 Gyr, reconfirming their results and favouring the crystallization-driven dynamo hypothesis. In summary, the incidence of magnetism in young (non-crystallized) WDs is about $\sim 10\%$, whereas this number raises to $\sim 30\%$ for old (crystallized) WDs.

Recent papers have studied the possibility that the convective motions induced by crystallization in the overlying liquid mantle are
efficient enough to explain the large intensity of the observed magnetic fields in the surface of WDs, ranging from $10^3$ to $10^9$ Gauss \citep{2022MNRAS.516L...1C,2022MNRAS.514.4111G,2024ApJ...961..197M}. Among these studies, \cite{2024ApJ...969...10C,2023ApJ...950...73F,2024ApJ...964L..15F} have shown that efficient convection could only take place at the onset of crystallization, and that thermohaline convection takes place during most of the crystallization process, thus demonstrating that only the dynamos driven at the onset of crystallization could account for the large intensity magnetic fields detected in WDs. Additionally, recent works by \cite{2024MNRAS.528.3153B,2024MNRAS.533L..13B} have shown that, due to the large electrical conductivity in the WD interior (and consequently, small magnetic diffusivity), even if the crystallization driven dynamos are able to account for the intensity of the observed magnetic field, this field cannot emerge to the surface immediately, taking from $\sim 1$ to $\sim 7$ Gyrs. 

In this paper we study a different mechanism that can account for the origin of magnetism in WDs and we compare it with the crystallization-driven dynamo hypothesis. Main sequence stars with masses $\gtrsim 1.1 M_\odot$ burn H through the CNO cycle thus developing convective cores, in which a magnetic dynamo is expected to take place. WDs more massive than $\sim 0.55M_\odot$ are the descendants of these main sequence stars with convective cores.
In the last decade, \cite{2015Sci...350..423F} proposed that the presence of strong magnetic fields in the core of red giant stars can alter the propagation of the gravity waves, decreasing the mode visibility in a phenomenon known as "magnetic greenhouse effect".
This way, asteroseismology has proven very strong magnetic fields (B $\gtrsim 10^5$ G ) that are trapped in the interior of many red giant branch stars, which are not detectable at their surfaces \citep{2023A&A...680A..26L,2023A&A...670L..16D,2022Natur.610...43L}. Indeed, \cite{2016Natur.529..364S} have studied the strength of dipolar oscillation modes in low- and intermediate-mass red giant stars, finding that strong core fields only occur in red giants more massive than 1.1 $M_\odot$. Also, they found that the occurrence rate is at least 50\% for stars with masses from 1.6 to 2.0$M_\odot$, indicating that powerful dynamos should take place in the { H-burning} convective cores of main sequence stars. \cite{2016ApJ...824...14C} proposed that this magnetic field generated during the main sequence convective core dynamo can survive all the evolutionary path into the WD phase. In this letter, we examine this possibility by 
estimating the breakout time in which this field, initially trapped in the stellar interior, should reach the stellar surface during the WD phase. Then, we compare those estimations with predictions for the emergence timescale of crystallization-driven dynamos and with the volume limited sample of magnetic WDs of \cite{2022ApJ...935L..12B}.

\section{Methods}
\label{Methods}

\subsection{Main sequence dynamo hypothesis}

Assuming an initial-to-final-mass relation \citep[e.g.][]{2008MNRAS.387.1693C,2018ApJ...866...21C}, 
the progenitors of WDs more massive than 
$\sim 0.55 M_\odot$ are main sequence stars with masses $\gtrsim 1.1 M_\odot$, which harbour convective cores during their main sequence phase, that could sustain dynamo magnetic fields \citep[see][]{2023SSRv..219...58K}. The strength of the generated magnetic fields is predicted to be in the range $10^4-10^6$G, by numerical magnetohydrodinamical (MHD) simulations \citep{2005ApJ...629..461B,2009ApJ...705.1000F,2024arXiv240918066H} and by estimations assuming equipartition between the magnetic energy
density and the kinetic energy density \citep{2016ApJ...824...14C}. Furthermore, the MHD simulations of \cite{2016ApJ...829...92A} showed that the field strength can be larger than  $10^6$G in the convective cores of 10$M_\odot$ B-type stars. 
Assuming  magnetic flux conservation, these values are compatible with the lower limits obtained for magnetic field strengths in the cores of red giant branch stars and in the surface of WD stars \citep{2016ApJ...824...14C,2016Natur.529..364S}.

	\begin{figure}
	\centering
	\includegraphics[clip,width=\columnwidth]{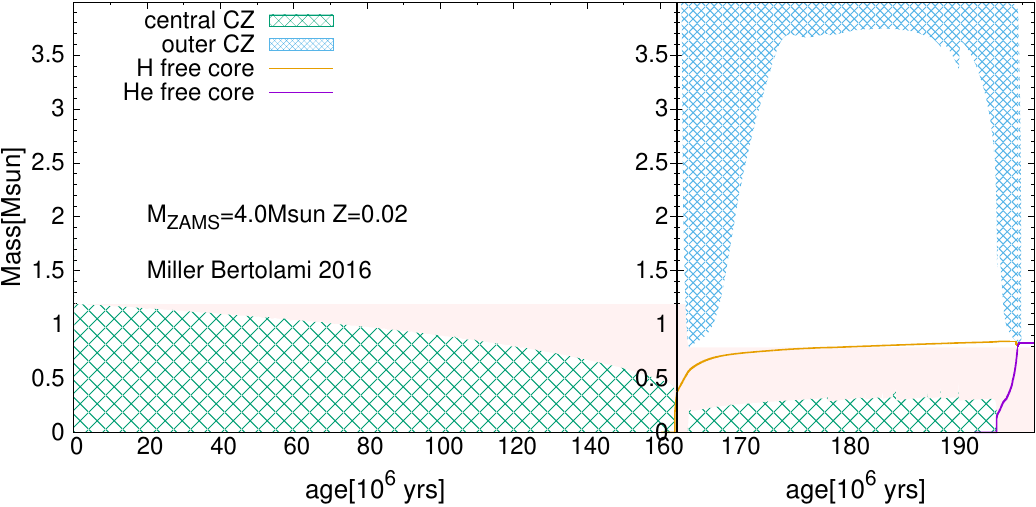}\\
    \caption{Kippenhahn diagram of a 4.0M$_\odot$ star with Z=0.02 from the main sequence to the end of the thermally pulsing AGB from \cite{2016A&A...588A..25M}. The hatched green and blue areas indicate the central and outer convective zones, respectively. The orange and purple lines indicate the outer limits of the H and He depleted regions, respectively. The pink shaded area is the region where the magnetic fields are expected to be buried. The intershell convective zones driven by the thermal pulses in the AGB are not visible in this plot. The final WD mass of this model is 0.83M$_\odot$.} 
    \label{fig:1}
\end{figure}

In Figure \ref{fig:1} we plot a Kippenhahn diagram from the main sequence to the planetary nebula phase of a 4.0M$_\odot$ star with Z=0.02, taken from \cite{2016A&A...588A..25M},  where the final WD mass is 0.83M$_\odot$. The green hatched area is the extension of the convective core, first during central H burning (left panel) and later during the central He burning (right panel). The hatched blue area shows the location of the outer convective zone. The deepest penetrations of the outer convective zone correspond to the first and second dredge-up, respectively. The pulse driven (or intershell) convective zones, which are short-lived and driven by the He shell flashes during the thermally pulsing AGB, are not visible in this plot due to
their short time scales. The boundaries of convective regions were set by the Schwarzschild criterion allowing for turbulent mixing beyond these boundaries \citep[see][for details on the overshooting treatment]{2016A&A...588A..25M}. 

In this paper we assume a magnetic field generated by dynamo action during the central H burning phase. Because the Ohmic diffusion timescales during the main sequence and red giant phases are too long \citep{2016ApJ...824...14C}, magnetic fields
present in the stellar core at main sequence are frozen in their Lagrangian mass coordinate. This way, during the main sequence (left panel), the magnetic field is confined to the pink shaded area, which denotes the maximum extension of the central convective zone. This assumption is probably not valid in areas of the star that become convective after the main sequence phase. 
Indeed, if the convective energy density, $\epsilon_{\rm con}$, exceeds the magnetic energy density, $\epsilon_{\rm mag}$, 
convection could distort the preexisting field into an unstable configuration. 
Conversely, if $\epsilon_{\rm con}<\epsilon_{\rm mag}$, the magnetic field can remain unchanged. We stress that it remains uncertain whether convection will disrupt preexisting stable magnetic field configurations \citep[see][for a thorough analysis]{2016ApJ...824...14C}. In this work, we simply assume that the magnetic field vanishes in the regions where the outer convective zone has penetrated. Therefore, the outer boundary of the magnetized region, that we call magnetic boundary (MB), is dictated by the maximum extent of the outer convective region, which can occur either during the first dredge-up or second dredge-up (in those stars in which a second dredge-up takes place). Nevertheless, in stars with masses lower than $\sim 2 M_\odot$, the outer convective zone does not reach the mass coordinates where the magnetic field was present and, for these stars, MB is determined by the maximum extent of the main sequence convective core.
Also, in these stars, a vigorous convection is developed in the core during core He flashes which will likely affect the magnetic field configuration. Furthermore, in such stars, the convective core during central He burning is larger than during central H burning, and the magnetic breakout times that we obtain for the WD descendants of these stars are larger than the Hubble timescale. Based on this information, we consider that in stars with initial masses lower than $\sim 2 M_\odot$ it is unlikely that the field can survive all the evolution and emerge in the WD phase.
Finally, we have ignored the fact that the pulse driven convective zones can alter the field configuration. However, we do not expect this choice to alter our main results, since in those WDs where the magnetic field can emerge to the surface within the Hubble timescale, the pulse driven convective regions are above the assumed magnetic field boundary.

\begin{figure}[ht!]
\includegraphics[
clip,width=\columnwidth]{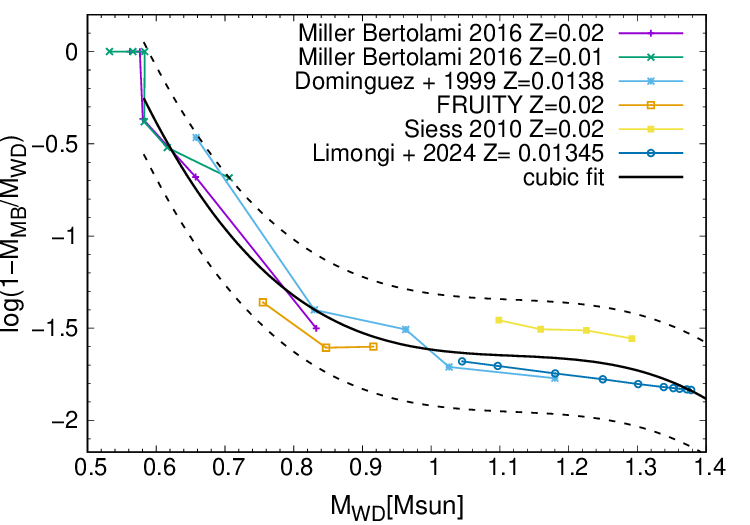}
    
\caption{{ Logarithm} of the outer mass fraction from the boundary where the magnetic field is expected to be trapped MB, 
in terms of the WD mass. We plot the theoretical predictions for CO WDs from \cite{2016A&A...588A..25M},\cite{1999ApJ...524..226D} and the Full-Network Repository of Updated Isotopic Tables \& Yields (F.R.U.I.T.Y.) Data Base \citep{2015ApJS..219...40C}, and for ONe WDs from \cite{2010A&A...512A..10S} and  \cite{2024ApJS..270...29L}. These models consider different metallicities spreading around the solar value.
A cubic polynomial fit to the data with $M_{\rm WD}>0.575M_\odot$ is shown using a solid black line. {A deviation of $\pm 0.30$ } from the cubic fit is shown using dotted black lines, which comprises all models considered in this work.}\label{fig:2}
\end{figure}

In Fig.  \ref{fig:2} we plot the boundaries of the magnetized regions, MB, that we obtained considering different stellar evolutionary models, i.e., the final outer boundary of the pink region in Fig. \ref{fig:1}. The x-axis is the WD mass, and the y-axis corresponds to the logarithm of 1 minus the mass coordinate at the magnetic boundary, MB, normalized to the WD mass. That is, the logarithm of the mass that the magnetic field has to diffuse through in order to reach the WD surface. It is important to stress that these values depend on the treatment of the convective boundaries and on the mass loss rates considered by each model. However, all sets of models predict that MB is closer to the surface for more massive WDs. Given the different values of the mass coordinate at MB that different stellar evolutionary codes and the different physical inputs predict, we have employed a cubic prescription to fit $\ln(1-M_{\rm MB}/M_{\rm WD})$ in terms of the WD mass (black solid line). We have allowed a deviation of { $\pm 0.30$dex} from the cubic fit (black dotted lines), which encompasses all the theoretical predictions from the different stellar evolutionary codes.

\subsection{Crystallization-driven dynamo hypothesis}

For the WD crystallization driven dynamos we consider MB as the outer boundary of the Rayleigh-Taylor unstable region induced by crystallization, as done in \cite{2024MNRAS.528.3153B}. We have employed the stellar evolutionary code {\tt LPCODE} \citep[see][for details]{2005A&A...435..631A} to calculate the WD evolutionary models, employing the phase diagram of \cite{2010PhRvL.104w1101H} to model CO crystallization. The outer boundary of the convective region is determined by imposing a positive carbon abundance gradient in the star $\nabla _ X= d \ln X /d \ln P \geq 0$ \citep[where, $X$ is the carbon abundance and $P$ is the pressure, see][for details]{2022MNRAS.511.5198C,2022MNRAS.516L...1C,2019A&A...625A..87C}. Note that the mass coordinate at MB for the crystallization dynamo moves outwards as evolution proceeds and the crystallized core grows in mass, unlike 
the MB for the main sequence dynamo that stays in the same mass coordinate during the entire WD phase.

\subsection{Magnetic diffusion and magnetic breakout times}

In order to obtain the magnetic breakout time ($t_{\rm br}$), we first estimate the diffusion time ($t_{\rm diff}$) that the magnetic field takes to diffuse from  MB to the WD surface, using \citep{2016ApJ...824...14C}:

\begin{equation}
\label{eq_diff}
t_{\rm diff}= \int_{r_{\rm MB}}^{R_{\rm WD}} \frac{{\rm d}(r-r_{\rm MB})^2}{\eta(r)}= \int_{r_{\rm MB}}^{R_{\rm WD}} \frac{2(r-r_{\rm MB}){\rm d} r}{\eta(r)}
\end{equation}
where $r_{\rm MB}$ is the radius that corresponds to MB, $R_{\rm WD}$ is the WD radius, and $\eta(r)$ is the magnetic diffusivity. $\eta(r)=c^2 /(4\pi \sigma(r) ) $, where $\sigma(r)$ is the electrical conductivity, which was obtained using equation (1) from \cite{2002MNRAS.333..589C} for the fully degenerate regime, the Spitzer formula for the non-degenerate regime (\cite{1962pfig.book.....S},  equation 19 in \cite{2002MNRAS.333..589C}), and an interpolation in between those regimes (i.e., when $0.1T_f<T<10T_f$, where $T_f$ is the Fermi temperature).
Note that $t_{\rm diff}$ at each time step is different, because $\sigma(r)$ and $r_{\rm MB}$ vary with time. We have calculated WD evolutionary models using {\tt LPCODE}, obtaining $\sigma(r)$, $r_{\rm MB}$ and $t_{\rm diff}$ at each time step. Once $t_{\rm diff}$ is calculated for each WD cooling time, $t_{\rm cool}$, we find $t_{\rm br}$ by calculating the earliest $t_{\rm br}$ when the equation:
\begin{equation}
t_{\rm br}= t_{\rm cool}+t_{\rm diff}(t_{\rm cool})
\end{equation}
is satisfied.

\section{Results}

\begin{figure}[ht!]
\includegraphics[
clip,width=\columnwidth]{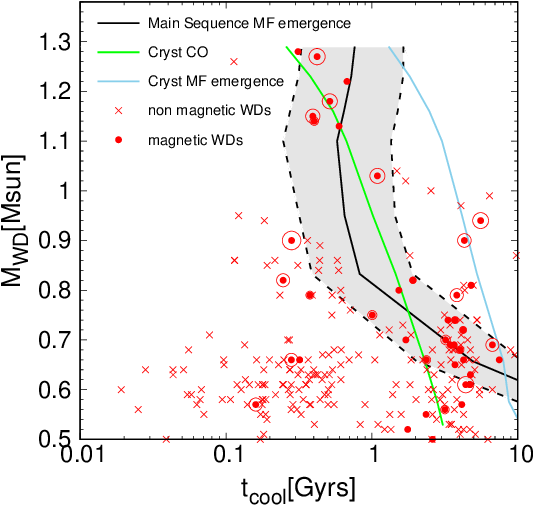}
\caption{Magnetic field breakout times for CO WDs. The crystallization onset is depicted using a green line and the magnetic field breakout time considering a crystallization-driven dynamo is plotted using a blue line. The magnetic breakout times considering a main sequence dynamo are shown using a black line for our cubic fit prescription, and the shaded are allows for a deviation of $\pm 0.3$ from it. The magnetic (non-magnetic) WDs from the spectropolarimetric surveys of \cite{2021MNRAS.507.5902B,2022ApJ...935L..12B} are plotted using filled red circles (red crosses), with surrounding circles whose radius are proportional to the magnetic field amplitude. These spectropolarimetric surveys are volume-limited up to 20pc for all WDs and  up to 40 pc for those WDs younger than 0.6 Gyrs.} 
    \label{fig:3}
\end{figure}

In Figure \ref{fig:3} we overlay the estimated magnetic field breakout times onto the distribution of WD masses vs. cooling ages for the sample of magnetic WDs  of 
\cite{2021MNRAS.507.5902B,2022ApJ...935L..12B}.
Our theoretical prediction for the magnetic field emergence for crystallization-driven dynamos in CO-core WDs (blue line) is in agreement with the results of \cite{2024MNRAS.528.3153B}.
We find that most of the magnetic WDs in this volume-limited sample cannot be explained by the crystallization-driven dynamo hypothesis, because they lie on the left of the magnetic field emergence line (blue line). Even if the crystallization-driven dynamo can sustain a  strong magnetic field, this field would be trapped in the WD interior for a very long period of time, ranging from $\sim 1 $Gyr in our most massive model (1.29$M_\odot$) to $\sim 7.3 $Gyrs in our 
least massive model (0.53$M_\odot$). 
On the contrary, if the main sequence dynamo magnetic field can survive trapped in the stellar interior all the way to the white phase, it can emerge to the surface accounting for many of the magnetic WDs in this sample. 
Depending on the parametrization of MB considered,  the magnetic field emergence should occur in a timescale longer than the Hubble time  for WDs with masses $\lesssim 0.6 M_\odot$. If we consider our fit for MB, allowing for a deviation of { $\pm$0.30 }dex, the main sequence dynamo magnetic field  emerges in the shaded area of Fig. \ref{fig:3}, accounting for most of the magnetic WDs in the sample. 
Nevertheless, the main sequence dynamo hypothesis cannot explain magnetic fields in WDs with masses $\lesssim 0.55 M_\odot$, because the main sequence progenitors of theses stars should have a radiative core. However, this hypothesis can explain the fact that magnetic field emergence occurs earlier the more massive the WDs, because MB is closer to the WD surface in these stars, regardless of the set of stellar models considered (see Fig. \ref{fig:2}). 
It is important to recall that the location of MB for the main sequence dynamo hypothesis is subject to enormous uncertainties in the modeling, as it depends on the treatment of convective boundaries, the mass loss rates, and the initial-to-final-mass relation. Therefore, we cannot give a certain prediction on the time for the main-sequence-dynamo field emergence, but we can say that it would take place within the WD phase for masses $\gtrsim 0.65 M_\odot$, and that it should occur earlier for more massive WDs. This hypothesis can explain the well known fact that magnetic WDs are in general more massive than non-magnetic WDs \citep{2020AdSpR..66.1025F,2020IAUS..357...60K}. 


\begin{figure}[ht!]
\includegraphics[
clip,width=\columnwidth]{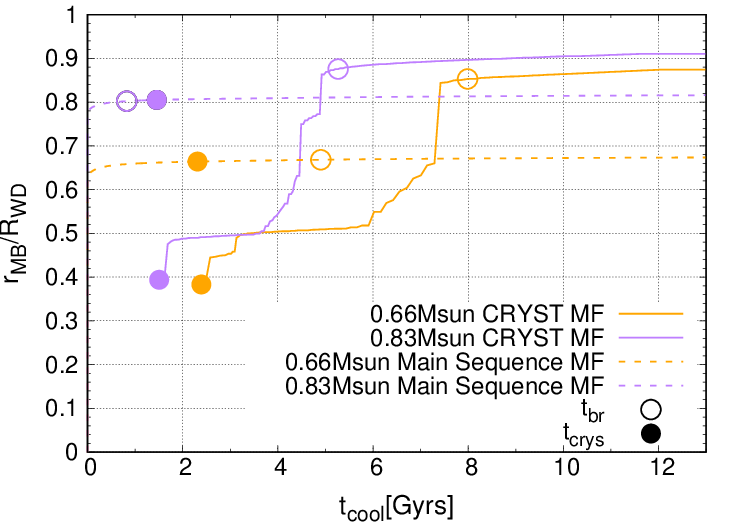}
\includegraphics[
clip,width=\columnwidth]{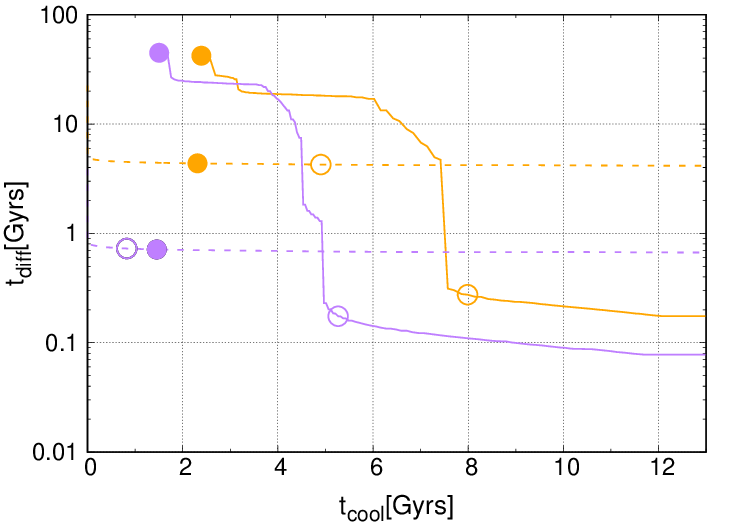}
\caption{Upper panel: Radius at the magnetic boundary, MB, normalized to the WD radius for our 0.66 and 0.83M$_\odot$ CO WD models. For the crystallization-driven dynamo models (solid lines), this radius is the outer boundary of the Rayleigh-Taylor unstable region induced by crystallization. For the main sequence dynamo models (dashed lines), this radius corresponds to the mass coordinate reached when the deepest penetration of the outer convective zone occurred in the WD progenitor evolution, either during the first or second dredge-up. The filled circle indicates the crystallization onset in our models, and the empty circle indicates the time when the magnetic field breaks out to the  WD surface. 
Lower panel: Magnetic diffusion time ($t_{\rm diff}$) calculated using Eq. \ref{eq_diff}
 for our 0.66 and 0.83M$_\odot$ WD models.}
     \label{fig:4}
\end{figure}

The radius (normalized to the WD radius) at MB and the magnetic diffusion times as a function of the WD cooling age are shown in the upper and lower panels of Figure \ref{fig:4}, respectively. The breakout time and the crystallization onset are marked using empty and filled circles, respectively. 
Although the normalized radius at MB stays nearly constant during the WD phase for the main sequence dynamo hypothesis, it increases drastically with the cooling time for the crystallization-dynamo hypothesis. This is because the outer convective boundary of the Rayleigh-Taylor unstable region moves outwards as the crystallized core grows. This boundary may also be subject to uncertainties in the chemical profile at the beginning of the WD phase \citep[see][for a through discussion]{2024MNRAS.528.3153B}. For the crystallization-dynamo hypothesis, we find diffusive times compatible with those of \cite{2024MNRAS.528.3153B}, although the breakout times we obtain are larger because these authors consider that the field starts to diffuse at the crystallization onset and we consider a time-dependent calculation. In both our 0.66 and 0.83 M$_\odot$ models, we obtain that the magnetic field breakout takes place earlier in the evolution when considering the main sequence dynamo hypothesis than if we consider the crystallization-driven dynamo hypothesis. We stress that the breakout timescale for crystallization-driven dynamos could be significantly longer than calculated here. As discussed by \cite{2024ApJ...964L..15F}, if crystallization is responsible for the strong magnetic fields observed in WDs, the dynamo must initiate at the onset of crystallization. However, due to the lack of detailed models for transport processes in 1D evolution codes, the extent of the convection zone is uncertain. The outer boundary of the convective region at the beginning of crystallization could lie anywhere between $\sim 0.1R_{WD}$ and $\sim 0.5R_{WD}$, depending on how convective mixing is implemented in the code (the lower limit assumes that mixing occurs according to the Ledoux criterion, whereas the upper limit, assumes mixing whenever the composition gradient is unstable). In the worst case scenario (Ledoux), the diffusion time for the magnetic field to travel from the outer boundary of the convective region to the WD surface would be much larger than the age of the Universe. In the best case scenario, it would be of the order of 12 Gyrs. Nevertheless strong magnetic fields can significantly enhance compositional mixing in thermohaline convection \citep[see, e.g.,][]{Harrington2019,Fraser2024}. This enhanced mixing could, in principle, push the outer boundary of the convective region closer to the WD's surface, shortening the timescale for magnetic field emergence.

\section{Conclusions}

WDs with masses larger than $\sim 0.55 M_\odot$ should have had convective cores in their main sequence progenitor life that can drive dynamo magnetic fields. This idea is supported by MHD simulations of convection in main sequence stars and by the strong magnetic fields inferred by asterosesismology in red giant branch stars. Although subsequent convective regions can disrupt this magnetic field, we have estimated that this field may survive trapped in the stellar interior all the way to the WD phase for stars with masses $M_{\rm ZAMS} \gtrsim 2 M_\odot$.  We have estimated the mass coordinate at which the magnetic field is supposed to be buried in WD stars, despite the large uncertainties in this boundary that the different stellar evolutionary codes and different physical inputs predict. We have calculated the time for the magnetic field to to diffuse from this boundary, MB, to the WD surface, finding the breakout times, $t_{\rm br}$, in which this field should emerge, for CO-core WDs in with masses in the range $0.58-1.29 M_\odot$. For comparison, we have calculated diffusive times and breakout times for the crystallization-driven magnetic fields, for CO-core WD in the mass range $0.53-1.29 M_\odot$. It is important to recall that, WDs with masses $\gtrsim 1.05 M_\odot$ should have ONe cores as predicted by the stellar evolutionary models of super AGB stars that we employed to obtain MB \citep{2010A&A...512A..10S,2024ApJS..270...29L}. However, a non-negligible fraction of WDs with masses $\gtrsim 1.05 M_\odot$ formed in single stellar evolution could have CO cores \citep{2021A&A...646A..30A}, thus supporting our assumption of CO-core for all WDs considered in this exploratory work. We defer the calculations for ONe WDs for a future work.

In agreement with previous results \citep{2024arXiv240601807C}, we find that, even if the crystallization-induced mixing can sustain a strong dynamo magnetic field, this hypothesis cannot explain the presence of magnetism in most of the magnetic WDs. This is because for the current mixing prescriptions implemented in evolution models, this field would be trapped in the WD interior for a very long time. Conversely, our calculations show that magnetic fields generated on the main sequence can emerge to the WD surface within the age of the Universe for WDs more massive than $\sim 0.65 M_\odot$. The more massive the WD, the earlier the field should emerge to the surface (in agreement with the crystallization-driven dynamos). The reason for this relies in the fact that, for stars more massive than $\sim 0.65 M_\odot$, the main sequence convective core comprised a mass larger than the final WD mass. However, we expect the outer convective zones penetrating during the first and second dredge-up to vanish the magnetic field in the outer layers of the star, thus burying the magnetic field in inner mass coordinates. This way, the outer stellar mass that the field has to diffuse through in order to reach the WD phase is smaller the larger the WD mass, and so is the magnetic breakout time. 

We cannot predict the precise location of the boundary of the trapped magnetic field from stellar models. This translates into an uncertainty of up to several Gyrs in the timing of the magnetic breakout. These uncertainties arise from factors such as the treatment of convective boundaries, mass loss rates, and the initial-to-final mass relation. Consequently, we cannot provide a definitive prediction for the emergence time of the main-sequence dynamo field. However, we can predict that, if a magnetic field is generated during the main sequence dynamo and it can survive trapped in the stellar interior all the evolution to the WD phase, it would emerge to the surface during the WD phase for masses $\gtrsim 0.65 M_\odot$. Also, it  will emerge to the surface earlier in more massive WDs. These results are consistent with the observations of the volume limited sample of magnetic WDs from \cite{2021MNRAS.507.5902B,2022ApJ...935L..12B}, and with the longstanding fact that magnetic WDs tend to be more massive than their non-magnetic counterparts. { It should be noted that magnetic fields generated as a result of stellar mergers can also explain the incidence of magnetism and the mass distribution of isolated magnetic white dwarfs, as shown by population synthesis studies \citep{2015MNRAS.447.1713B,2012ApJ...749...25G,2008MNRAS.387..897T}.}

As a significant number of strongly magnetic WDs are found in close binaries, the predictions of any scenario for the origin of WD magnetic fields should be confronted with observations of magnetic WDs in close binaries. 
\citet{schreiberetal21-1} recently suggested an evolutionary sequence based on the crystallization driven dynamo that for the first time explains the existence and relative numbers of magnetic WDs in cataclysmic variables, detached WD binaries with strongly magnetic WDs \citep{parsonsetal21-1}, and radio pulsing WD binary stars \citep{marshetal16-1}.
Assuming instead of the crystallization driven dynamo the main sequence dynamo we propose here, this evolutionary sequence remains plausible. As long as strong magnetic fields appear in WDs in cataclysmic variables, angular momentum transfer from the WD to the orbit, a subsequent detached phase as well as synchronization are predicted.

In any case, it seems likely that the main sequence dynamo cannot explain the magnetic field in all WDs, and there is likely a variety of channels for the occurrence of WD magnetism. In particular, it cannot account for magnetism in WDs that descend from main sequence stars that hold radiative cores. Nevertheless, this scenario can account for the young massive magnetic WDs observed in globular clusters which cannot be explained by the merger or crystallization hypotheses \citep{2020ApJ...901L..14C}. 
It is important to stress that there is no clear evidence that the magnetic field is trapped in the stellar interior of red giant stars nor clump stars more massive than $\sim 2 M_\odot$ \citep{2024MNRAS.528.7397C}. 
In order to adequately test the capability of main sequence dynamos to explain WD magnetism, a through analysis of the field evolution through all stages of stellar evolution should be carried out. Furthermore, a better estimation of time of emergence and the predicted surface magnetic field strength can be obtained by solving the induction equation as in \cite{2024arXiv240601807C}. Finally, we want to emphasize the capabilities of magnetic WDs to understand the efficiency of main sequence dynamos, in a similar manner to asteroseismology of red giant stars. 
Indeed, the study of
magnetic WDs can help to improve our understanding of stellar dynamos and stellar evolution.

\begin{acknowledgements}
 MC acknowledges
grant RYC2021-032721-I, funded by MCIN/AEI/10.13039/501100011033 and by the European Union NextGenerationEU/PRTR. J.R.F. is supported by NASA Solar System Workings grant 80NSSC24K0927. 
MRS thanks for support from FONDECYT (grant number 1221059). 
  This work was partially supported by the AGAUR/Generalitat de Catalunya grant SGR-386/2021, by the Spanish MINECO grant PID2020-117252GB-I00 and by  PID2021-123110NB-I00  financed by
MCIN/AEI/10.13039/501100011033/FEDER,  UE.
 This research was supported by the Munich Institute for Astro-, Particle and BioPhysics (MIAPbP), which is funded by the Deutsche Forschungsgemeinschaft (DFG, German Research Foundation) under Germany´s Excellence Strategy – EXC-2094 – 390783311. MC acknowledges Jim Fuller, Lilia Ferrario,  Stefano Bagnulo and Ilaria Caiazzo for useful discussions. { The authors acknowledge the anonymous referee, whose comments have helped to improve the manuscript.}
\end{acknowledgements}

\bibliographystyle{aa} 
\bibliography{lowZ}

\begin{thebibliography}{56}
\expandafter\ifx\csname natexlab\endcsname\relax\def\natexlab#1{#1}\fi

\bibitem[{{Althaus} {et~al.}(2015){Althaus}, {Camisassa}, {Miller Bertolami},
  {C{\'o}rsico}, \& {Garc{\'{\i}}a-Berro}}]{2015A&A...576A...9A}
{Althaus}, L.~G., {Camisassa}, M.~E., {Miller Bertolami}, M.~M., {C{\'o}rsico},
  A.~H., \& {Garc{\'{\i}}a-Berro}, E. 2015, \aap, 576, A9

\bibitem[{{Althaus} {et~al.}(2010){Althaus}, {C{\'o}rsico}, {Isern}, \&
  {Garc{\'{\i}}a-Berro}}]{2010A&ARv..18..471A}
{Althaus}, L.~G., {C{\'o}rsico}, A.~H., {Isern}, J., \& {Garc{\'{\i}}a-Berro},
  E. 2010, \aapr, 18, 471

\bibitem[{{Althaus} {et~al.}(2021){Althaus}, {Gil-Pons}, {C{\'o}rsico}, {Miller
  Bertolami}, {De Ger{\'o}nimo}, {Camisassa}, {Torres}, {Gutierrez}, \&
  {Rebassa-Mansergas}}]{2021A&A...646A..30A}
{Althaus}, L.~G., {Gil-Pons}, P., {C{\'o}rsico}, A.~H., {et~al.} 2021, \aap,
  646, A30

\bibitem[{{Althaus} {et~al.}(2005){Althaus}, {Serenelli}, {Panei},
  {C{\'o}rsico}, {Garc{\'{\i}}a-Berro}, \&
  {Sc{\'o}ccola}}]{2005A&A...435..631A}
{Althaus}, L.~G., {Serenelli}, A.~M., {Panei}, J.~A., {et~al.} 2005, A\&A, 435,
  631

\bibitem[{{Angel} \& {Landstreet}(1970)}]{1970ApJ...160L.147A}
{Angel}, J.~R.~P. \& {Landstreet}, J.~D. 1970, \apjl, 160, L147

\bibitem[{{Augustson} {et~al.}(2016){Augustson}, {Brun}, \&
  {Toomre}}]{2016ApJ...829...92A}
{Augustson}, K.~C., {Brun}, A.~S., \& {Toomre}, J. 2016, \apj, 829, 92

\bibitem[{{Bagnulo} \& {Landstreet}(2021)}]{2021MNRAS.507.5902B}
{Bagnulo}, S. \& {Landstreet}, J.~D. 2021, \mnras, 507, 5902

\bibitem[{{Bagnulo} \& {Landstreet}(2022)}]{2022ApJ...935L..12B}
---. 2022, \apjl, 935, L12

\bibitem[{{Blatman} \& {Ginzburg}(2024{\natexlab{a}})}]{2024MNRAS.528.3153B}
{Blatman}, D. \& {Ginzburg}, S. 2024{\natexlab{a}}, \mnras, 528, 3153

\bibitem[{{Blatman} \& {Ginzburg}(2024{\natexlab{b}})}]{2024MNRAS.533L..13B}
---. 2024{\natexlab{b}}, \mnras, 533, L13

\bibitem[{{Briggs} {et~al.}(2015){Briggs}, {Ferrario}, {Tout},
  {Wickramasinghe}, \& {Hurley}}]{2015MNRAS.447.1713B}
{Briggs}, G.~P., {Ferrario}, L., {Tout}, C.~A., {Wickramasinghe}, D.~T., \&
  {Hurley}, J.~R. 2015, \mnras, 447, 1713

\bibitem[{{Brun} {et~al.}(2005){Brun}, {Browning}, \&
  {Toomre}}]{2005ApJ...629..461B}
{Brun}, A.~S., {Browning}, M.~K., \& {Toomre}, J. 2005, \apj, 629, 461

\bibitem[{{Caiazzo} {et~al.}(2020){Caiazzo}, {Heyl}, {Richer}, {Cummings},
  {Fleury}, {Hegarty}, {Kalirai}, {Kerr}, {Thiele}, {Tremblay}, \&
  {Villanueva}}]{2020ApJ...901L..14C}
{Caiazzo}, I., {Heyl}, J., {Richer}, H., {et~al.} 2020, \apjl, 901, L14

\bibitem[{{Camisassa} {et~al.}(2019){Camisassa}, {Althaus}, {C{\'o}rsico}, {De
  Ger{\'o}nimo}, {Miller Bertolami}, {Novarino}, {Rohrmann}, {Wachlin}, \&
  {Garc{\'\i}a-Berro}}]{2019A&A...625A..87C}
{Camisassa}, M.~E., {Althaus}, L.~G., {C{\'o}rsico}, A.~H., {et~al.} 2019,
  \aap, 625, A87

\bibitem[{{Camisassa} {et~al.}(2022{\natexlab{a}}){Camisassa}, {Althaus},
  {Koester}, {Torres}, {Gil-Pons}, \& {C{\'o}rsico}}]{2022MNRAS.511.5198C}
{Camisassa}, M.~E., {Althaus}, L.~G., {Koester}, D., {et~al.}
  2022{\natexlab{a}}, \mnras, 511, 5198

\bibitem[{{Camisassa} {et~al.}(2022{\natexlab{b}}){Camisassa}, {Raddi},
  {Althaus}, {Isern}, {Rebassa-Mansergas}, {Torres}, {C{\'o}rsico}, \&
  {Korre}}]{2022MNRAS.516L...1C}
{Camisassa}, M.~E., {Raddi}, R., {Althaus}, L.~G., {et~al.} 2022{\natexlab{b}},
  \mnras, 516, L1

\bibitem[{{Cantiello} {et~al.}(2016){Cantiello}, {Fuller}, \&
  {Bildsten}}]{2016ApJ...824...14C}
{Cantiello}, M., {Fuller}, J., \& {Bildsten}, L. 2016, \apj, 824, 14

\bibitem[{{Castro-Tapia} {et~al.}(2024{\natexlab{a}}){Castro-Tapia}, {Cumming},
  \& {Fuentes}}]{2024ApJ...969...10C}
{Castro-Tapia}, M., {Cumming}, A., \& {Fuentes}, J.~R. 2024{\natexlab{a}},
  \apj, 969, 10

\bibitem[{{Castro-Tapia} {et~al.}(2024{\natexlab{b}}){Castro-Tapia}, {Zhang},
  \& {Cumming}}]{2024arXiv240601807C}
{Castro-Tapia}, M., {Zhang}, S., \& {Cumming}, A. 2024{\natexlab{b}}, arXiv
  e-prints, arXiv:2406.01807

\bibitem[{{Catal{\'a}n} {et~al.}(2008){Catal{\'a}n}, {Isern},
  {Garc{\'\i}a-Berro}, \& {Ribas}}]{2008MNRAS.387.1693C}
{Catal{\'a}n}, S., {Isern}, J., {Garc{\'\i}a-Berro}, E., \& {Ribas}, I. 2008,
  \mnras, 387, 1693

\bibitem[{{Crawford} {et~al.}(2024){Crawford}, {Bedding}, {Li}, {Stello},
  {Huber}, {Yu}, {Sreenivas}, {Li}, \& {Kerrison}}]{2024MNRAS.528.7397C}
{Crawford}, C.~L., {Bedding}, T.~R., {Li}, Y., {et~al.} 2024, \mnras, 528, 7397

\bibitem[{{Cristallo} {et~al.}(2015){Cristallo}, {Straniero}, {Piersanti}, \&
  {Gobrecht}}]{2015ApJS..219...40C}
{Cristallo}, S., {Straniero}, O., {Piersanti}, L., \& {Gobrecht}, D. 2015,
  \apjs, 219, 40

\bibitem[{{Cumming}(2002)}]{2002MNRAS.333..589C}
{Cumming}, A. 2002, \mnras, 333, 589

\bibitem[{{Cummings} {et~al.}(2018){Cummings}, {Kalirai}, {Tremblay},
  {Ramirez-Ruiz}, \& {Choi}}]{2018ApJ...866...21C}
{Cummings}, J.~D., {Kalirai}, J.~S., {Tremblay}, P.~E., {Ramirez-Ruiz}, E., \&
  {Choi}, J. 2018, \apj, 866, 21

\bibitem[{{De Ger{\'o}nimo} {et~al.}(2019){De Ger{\'o}nimo}, {Battich}, {Miller
  Bertolami}, {Althaus}, \& {C{\'o}rsico}}]{2019A&A...630A.100D}
{De Ger{\'o}nimo}, F.~C., {Battich}, T., {Miller Bertolami}, M.~M., {Althaus},
  L.~G., \& {C{\'o}rsico}, A.~H. 2019, \aap, 630, A100

\bibitem[{{Deheuvels} {et~al.}(2023){Deheuvels}, {Li}, {Ballot}, \&
  {Ligni{\`e}res}}]{2023A&A...670L..16D}
{Deheuvels}, S., {Li}, G., {Ballot}, J., \& {Ligni{\`e}res}, F. 2023, \aap,
  670, L16

\bibitem[{{Dominguez} {et~al.}(1999){Dominguez}, {Chieffi}, {Limongi}, \&
  {Straniero}}]{1999ApJ...524..226D}
{Dominguez}, I., {Chieffi}, A., {Limongi}, M., \& {Straniero}, O. 1999, \apj,
  524, 226

\bibitem[{{Featherstone} {et~al.}(2009){Featherstone}, {Browning}, {Brun}, \&
  {Toomre}}]{2009ApJ...705.1000F}
{Featherstone}, N.~A., {Browning}, M.~K., {Brun}, A.~S., \& {Toomre}, J. 2009,
  \apj, 705, 1000

\bibitem[{{Ferrario} {et~al.}(2015){Ferrario}, {de Martino}, \&
  {G{\"a}nsicke}}]{2015SSRv..191..111F}
{Ferrario}, L., {de Martino}, D., \& {G{\"a}nsicke}, B.~T. 2015, \ssr, 191, 111

\bibitem[{{Ferrario} {et~al.}(2020){Ferrario}, {Wickramasinghe}, \&
  {Kawka}}]{2020AdSpR..66.1025F}
{Ferrario}, L., {Wickramasinghe}, D., \& {Kawka}, A. 2020, Advances in Space
  Research, 66, 1025

\bibitem[{{Fraser} {et~al.}(2024){Fraser}, {Reifenstein}, \&
  {Garaud}}]{Fraser2024}
{Fraser}, A.~E., {Reifenstein}, S.~A., \& {Garaud}, P. 2024, \apj, 964, 184

\bibitem[{{Fuentes} {et~al.}(2024){Fuentes}, {Castro-Tapia}, \&
  {Cumming}}]{2024ApJ...964L..15F}
{Fuentes}, J.~R., {Castro-Tapia}, M., \& {Cumming}, A. 2024, \apjl, 964, L15

\bibitem[{{Fuentes} {et~al.}(2023){Fuentes}, {Cumming}, {Castro-Tapia}, \&
  {Anders}}]{2023ApJ...950...73F}
{Fuentes}, J.~R., {Cumming}, A., {Castro-Tapia}, M., \& {Anders}, E.~H. 2023,
  \apj, 950, 73

\bibitem[{{Fuller} {et~al.}(2015){Fuller}, {Cantiello}, {Stello}, {Garcia}, \&
  {Bildsten}}]{2015Sci...350..423F}
{Fuller}, J., {Cantiello}, M., {Stello}, D., {Garcia}, R.~A., \& {Bildsten}, L.
  2015, Science, 350, 423

\bibitem[{{Garc{\'{\i}}a-Berro} {et~al.}(2012){Garc{\'{\i}}a-Berro},
  {Lor{\'e}n-Aguilar}, {Aznar-Sigu{\'a}n}, {Torres}, {Camacho}, {Althaus},
  {C{\'o}rsico}, {K{\"u}lebi}, \& {Isern}}]{2012ApJ...749...25G}
{Garc{\'{\i}}a-Berro}, E., {Lor{\'e}n-Aguilar}, P., {Aznar-Sigu{\'a}n}, G.,
  {et~al.} 2012, \apj, 749, 25

\bibitem[{{Ginzburg} {et~al.}(2022){Ginzburg}, {Fuller}, {Kawka}, \&
  {Caiazzo}}]{2022MNRAS.514.4111G}
{Ginzburg}, S., {Fuller}, J., {Kawka}, A., \& {Caiazzo}, I. 2022, \mnras, 514,
  4111

\bibitem[{{Harrington} \& {Garaud}(2019)}]{Harrington2019}
{Harrington}, P.~Z. \& {Garaud}, P. 2019, \apjl, 870, L5

\bibitem[{{Hidalgo} {et~al.}(2024){Hidalgo}, {K{\"a}pyl{\"a}}, {Schleicher},
  {Ortiz-Rodr{\'\i}guez}, \& {Navarrete}}]{2024arXiv240918066H}
{Hidalgo}, J.~P., {K{\"a}pyl{\"a}}, P.~J., {Schleicher}, D.~R.~G.,
  {Ortiz-Rodr{\'\i}guez}, C.~A., \& {Navarrete}, F.~H. 2024, arXiv e-prints,
  arXiv:2409.18066

\bibitem[{{Horowitz} {et~al.}(2010){Horowitz}, {Schneider}, \&
  {Berry}}]{2010PhRvL.104w1101H}
{Horowitz}, C.~J., {Schneider}, A.~S., \& {Berry}, D.~K. 2010, Physical Review
  Letters, 104, 231101

\bibitem[{{Isern} {et~al.}(2017){Isern}, {Garc{\'\i}a-Berro}, {K{\"u}lebi}, \&
  {Lor{\'e}n-Aguilar}}]{2017ApJ...836L..28I}
{Isern}, J., {Garc{\'\i}a-Berro}, E., {K{\"u}lebi}, B., \& {Lor{\'e}n-Aguilar},
  P. 2017, \apjl, 836, L28

\bibitem[{{K{\"a}pyl{\"a}} {et~al.}(2023){K{\"a}pyl{\"a}}, {Browning}, {Brun},
  {Guerrero}, \& {Warnecke}}]{2023SSRv..219...58K}
{K{\"a}pyl{\"a}}, P.~J., {Browning}, M.~K., {Brun}, A.~S., {Guerrero}, G., \&
  {Warnecke}, J. 2023, \ssr, 219, 58

\bibitem[{{Kawka}(2020)}]{2020IAUS..357...60K}
{Kawka}, A. 2020, in IAU Symposium, Vol. 357, White Dwarfs as Probes of
  Fundamental Physics: Tracers of Planetary, Stellar and Galactic Evolution,
  ed. M.~A. {Barstow}, S.~J. {Kleinman}, J.~L. {Provencal}, \& L.~{Ferrario},
  60--74

\bibitem[{{Kawka} {et~al.}(2007){Kawka}, {Vennes}, {Schmidt}, {Wickramasinghe},
  \& {Koch}}]{2007ApJ...654..499K}
{Kawka}, A., {Vennes}, S., {Schmidt}, G.~D., {Wickramasinghe}, D.~T., \&
  {Koch}, R. 2007, \apj, 654, 499

\bibitem[{{Kemp} {et~al.}(1970){Kemp}, {Swedlund}, {Landstreet}, \&
  {Angel}}]{1970ApJ...161L..77K}
{Kemp}, J.~C., {Swedlund}, J.~B., {Landstreet}, J.~D., \& {Angel}, J.~R.~P.
  1970, \apjl, 161, L77

\bibitem[{{Li} {et~al.}(2022){Li}, {Deheuvels}, {Ballot}, \&
  {Ligni{\`e}res}}]{2022Natur.610...43L}
{Li}, G., {Deheuvels}, S., {Ballot}, J., \& {Ligni{\`e}res}, F. 2022, \nat,
  610, 43

\bibitem[{{Li} {et~al.}(2023){Li}, {Deheuvels}, {Li}, {Ballot}, \&
  {Ligni{\`e}res}}]{2023A&A...680A..26L}
{Li}, G., {Deheuvels}, S., {Li}, T., {Ballot}, J., \& {Ligni{\`e}res}, F. 2023,
  \aap, 680, A26

\bibitem[{{Limongi} {et~al.}(2024){Limongi}, {Roberti}, {Chieffi}, \&
  {Nomoto}}]{2024ApJS..270...29L}
{Limongi}, M., {Roberti}, L., {Chieffi}, A., \& {Nomoto}, K. 2024, \apjs, 270,
  29

\bibitem[{{Marsh} {et~al.}(2016){Marsh}, {G{\"a}nsicke}, {H{\"u}mmerich},
  {Hambsch}, {Bernhard}, {Lloyd}, {Breedt}, {Stanway}, {Steeghs}, {Parsons},
  {Toloza}, {Schreiber}, {Jonker}, {van Roestel}, {Kupfer}, {Pala}, {Dhillon},
  {Hardy}, {Littlefair}, {Aungwerojwit}, {Arjyotha}, {Koester}, {Bochinski},
  {Haswell}, {Frank}, \& {Wheatley}}]{marshetal16-1}
{Marsh}, T.~R., {G{\"a}nsicke}, B.~T., {H{\"u}mmerich}, S., {et~al.} 2016,
  \nat, 537, 374

\bibitem[{{Miller Bertolami}(2016)}]{2016A&A...588A..25M}
{Miller Bertolami}, M.~M. 2016, \aap, 588, A25

\bibitem[{{Montgomery} \& {Dunlap}(2024)}]{2024ApJ...961..197M}
{Montgomery}, M.~H. \& {Dunlap}, B.~H. 2024, \apj, 961, 197

\bibitem[{{Parsons} {et~al.}(2021){Parsons}, {G{\"a}nsicke}, {Schreiber},
  {Marsh}, {Ashley}, {Breedt}, {Littlefair}, \& {Meusinger}}]{parsonsetal21-1}
{Parsons}, S.~G., {G{\"a}nsicke}, B.~T., {Schreiber}, M.~R., {et~al.} 2021,
  \mnras, 502, 4305

\bibitem[{{Schreiber} {et~al.}(2021){Schreiber}, {Belloni}, {G{\"a}nsicke},
  {Parsons}, \& {Zorotovic}}]{schreiberetal21-1}
{Schreiber}, M.~R., {Belloni}, D., {G{\"a}nsicke}, B.~T., {Parsons}, S.~G., \&
  {Zorotovic}, M. 2021, Nature Astronomy, 5, 648

\bibitem[{{Siess}(2010)}]{2010A&A...512A..10S}
{Siess}, L. 2010, \aap, 512, A10

\bibitem[{{Spitzer}(1962)}]{1962pfig.book.....S}
{Spitzer}, L. 1962, {Physics of Fully Ionized Gases}

\bibitem[{{Stello} {et~al.}(2016){Stello}, {Cantiello}, {Fuller}, {Huber},
  {Garc{\'\i}a}, {Bedding}, {Bildsten}, \& {Silva
  Aguirre}}]{2016Natur.529..364S}
{Stello}, D., {Cantiello}, M., {Fuller}, J., {et~al.} 2016, \nat, 529, 364

\bibitem[{{Tout} {et~al.}(2008){Tout}, {Wickramasinghe}, {Liebert}, {Ferrario},
  \& {Pringle}}]{2008MNRAS.387..897T}
{Tout}, C.~A., {Wickramasinghe}, D.~T., {Liebert}, J., {Ferrario}, L., \&
  {Pringle}, J.~E. 2008, \mnras, 387, 897

\end{thebibliography}
 
\end{document}